\documentclass[checkout,aps,pra,twocolumn]{revtex4}

\newcommand{\ba}{\begin{eqnarray}}
\newcommand{\ea}{\end{eqnarray}}

\usepackage{graphicx,color}

\newcommand{\ce}{$\rm C_{60}$}
\newcommand{\cm}{$\rm C_{60}$ }

\newcommand{\be}{\begin{equation}}
\newcommand{\ee}{\end{equation}}

\newcommand{\et}{{\it et al. }}

\def\prl{{ Phys. Rev. Lett. }}

\def\prb{{ Phys. Rev. B }}

\begin{document}

%\title{Transient electromagnetically induced transparency in
%fullerenes }

%\title{Laser-induced coherent population trapping in \ce}

%\title{ Mysterious anistropic velocity-map images in ${\bf C_{60}}$
%  are the precursor of 1f superatom molecular orbital}

%\title{Manifestation of ${\bf C_{60}}$ 1f-superatom
%  molecular orbital in mysterious velocity-map images through four
%  800-nm photon excitation }

%\title{ Mysterious anistropic velocity-map images in ${\bf C_{60}}$
%  are the precursor of 1f superatom molecular orbital}

%\title{Detection of the ${\bf C_{60}}$ 1f-superatom molecular orbital
%  through four-photon (800 nm) excitations }

%\title{${\bf C_{60}}$'s 1f-superatom molecular orbital emerges from
%  mysterious velocity-map images through four-800-nm photon
%  excitations }

\title{ Superintermolecular orbitals in the C$_{60}$/pentacene complex}

 \author{G. P. Zhang,$^{1,*}$ A. Gardner,$^1$ T. Latta,$^1$
   K. Drake$^1$ and Y. H. Bai$^2$}

 \affiliation{$^1$Department of Physics, Indiana State University, Terre
   Haute, Indiana 47809, USA}

\affiliation{$^2$Office of Information Technology, Indiana State
  University, Terre Haute, IN 47809, USA}

 \date{\today}

\begin{abstract}
{We report a group of unusually big molecular orbitals in the
  \ce/pentacene complex.  Our first-principles density functional
  calculation shows that these orbitals are very delocalized and cover
  both \cm and pentacene, which we call superintermolecular orbitals
  or SIMOs.  Their spatial extension can reach 1 nm or larger.
  Optically, SIMOs are dark. Different from ordinary unoccupied
  molecular orbitals, SIMOs have a very weak Coulomb and exchange
  interaction. Their energy levels are very similar to the native
  superatomic molecular orbitals in \ce, and can be approximately
  characterized by orbital angular momentum quantum numbers.  They
  have a distinctive spatial preference.  These features fit the key
  characters of charge-generation states that channel initially-bound
  electrons and holes into free charge carriers. Thus, our finding is
  important for \ce/pentacene photovoltaics. }
\end{abstract}
\maketitle

\section{Introduction}

Organic solar cells are flexible, stretchable and possibly
wearable. If they could be integrated into our clothing, they would
power our portable phones and computers.  \ce/pentacene solar cells
are a prime example in organic photovoltaics \cite{bredas}, where
pentacene (Pc) serves as an electron donor and \cm as an electron
acceptor. When light strikes on pentacene, a complex singlet is formed
and subsequently is split into two optically inactive triplets, or
singlet fission \cite{chan,thor,zimmerman,wilson,sanders}. Such a
unique feature, where one single photon creates two triplets, greatly
improves the quantum efficiency of charge
photogeneration \cite{congreve}. 

But the high quantum efficiency is only the first step for the
photovoltaic cell \cite{congreve}. What is more important is the states
that channel initially-bound electrons and holes into free charge
carriers. Bakulin \et \cite{bakulin} showed that the formation of
delocalized states facilitates photoconversion. In 2014, Gelinas \et
\cite{gelinas} suggested that a rapid (40 fs) charge separation
proceeds through delocalized $\pi$-electron states in ordered regions
of the fullerene and acceptor material. Chen \et \cite{chen} also
found that charge photogeneration occurs predominantly via those
delocalized hot exciton states. Paraecatti and Banerji \cite{para}
more directly pointed out that exciton delocalization provides an
efficient charge separation pathway. These prior studies established
beyond any doubt the important of delocalized states, but what are
these delocalized channel states \cite{savoie,savoie2}? To this end,
there has been no obvious answer. This is the focus of our study.

In this paper, we carry out an extensive first-principles density
functional calculation to show that there are a group of unusually
larger superintermolecular orbitals (SIMOs) in \ce/Pc complex that
bridge both \cm and pentacene.  We employ a real grid mesh method so
we can treat both ordinary molecular orbitals and SIMOs on an equal
footing. We find that energetically, SIMOS are close to native
superatom molecular orbitals in \ce, but spatially SIMOs are much
larger, with spatial extension over 1 nm.  They are optically
silent. By computing over 3000 Coulomb and exchange integrals, we find
that both Coulomb and exchange interactions among SIMOs are in general
much smaller than those among ordinary molecular orbitals, a necessary
condition to allow initially bound electrons and holes to dissociate
into free charge carriers. Interestingly, regardless of edge-on and
face-one geometries, SIMOs retain their original shapes.  These
features strongly suggest that they are good candidates for those
channel states in \ce/Pc solar cells.

The rest of the paper is arranged as follows. In Sec. II, we present
our theoretical formalism and the details of our first-principles
calculation.  Section III is devoted to the results and discussion.
We conclude this paper in Sec. IV.  An appendix at end provides
additional details about our hybrid MPI/OpenMP parallel
implementation.

\section{Method}

Our calculation is based on the first-principles density functional
code Octopus \cite{octopus} which employs the pseudopotential method
and the real grid mesh in real space, and has an important advantage
that it treats localized and delocalized states on an equal footing.
To start with, we solve the Kohn-Sham (KS) equation in atomic units,
\be \left [-\frac{1}{2}\nabla^2+V_{eff} ({\bf r})\right ] \phi_i ({\bf
  r})=E_i\phi_i ({\bf r})\label{ks} \ee where $\phi_i ({\bf r})$ is
the Kohn-Sham wavefunction and $E_i$ is the eigenvalue of state $i$.
The first term on the left hand side of Eq. (\ref{ks}) is the kinetic
energy operator.  The effective potential ($V_{eff}$) consists of the
electron-nuclei interaction, the Hartree potential (due to the
electron-electron Coulomb interaction), and exchange-correlation
interactions, \be V_{eff} ({\bf r})=v({\bf r})+ \int d{\bf r'}
\frac{\rho({\bf r'})}{|{\bf r}-{\bf r'}|} +V_{xc}({\bf r}) \ee where
the exchange-correlation potential $V_{xc}$ is $\delta
E_{xc}[\rho]/\delta \rho({\bf r})$, taking the form of the local
density approximation (LDA). We find that LDA is sufficient for our
purpose, and using GGA raises the energy by 0.5 eV \cite{ijmp2015}.
The new charge density is computed by summing over all the occupied
orbitals ($N_{occ}$), \be \rho({\bf r})=\sum_{i=1}^{N_{occ}}|\phi_i
({\bf r})|^2. \ee The next iteration starts. This process repeats
itself until the charge density converges.  With the converged
wavefunction, we then compute the Coulomb and exchange integrals using
National Energy Research Scientific Computing Center machines.
However, these integrals over six degrees of freedom are extremely
time consuming, with so many mesh grid points (see the appendix for
details).  We employ the submatrix technique where we compute the
action of the Coulomb term on the states $n$ and $m$ and then multiply
two additional wavefunctions on the above results.  We develop a
hybrid MPI/OpenMP code that breaks the integral into segments and
distribute them to processors and nodes.  And finally, the master node
sums all the results up. This speeds up our calculation greatly.

We use the normal conserving pseudopotential developed by Troullier
and Martins \cite{tm}.  Our simulation box is a cylinder. The radius
of the cylinder is $r=30~\rm \AA$ and the length is 80 $\rm \AA$. The
grid mesh is $m=0.22~\rm \AA$ and the total number of grid mesh points
is 22814131.  We have checked the convergence with the grid mesh and
find that our results converge well.  $\rm C_{60}$/pentacene complex
has 342 valence electrons, so 171 orbitals are doubly occupied. To
obtain those unoccupied states, we add 129 extra states (in Octopus,
the command is ExtraStates=129), so we have eigenstates all the way up
to 300.  This covers the entire spectrum that is of interest to us.
The threshold for the charge density convergence is set to $10^{-4}$,
and the threshold for the absolute energy convergence is set to
$5\times 10^{-7}$ eV. All the Octopus calculations are run on our
university Silicon cluster, where each computing node has dual Intel
Xeon E5-2680 v2 CPUs with 2.80GHZ.  Each CPU has 10 cores and cache
size of 25 MB. The total memory for each node is 132 GB. The entire
calculation needs 80 GB memory and takes nearly two months to finish.
After the calculation is finished, we export the wavefunctions from
state number 97 up to 300 in two different formats, one for Xcrysden
rendering of orbital images and the other in the Cartesian format. The
latter is the actual wavefunctions in the three-dimensional space
$\phi(x,y,z)$. These wavefunctions are extremely convenient for
calculating other properties of interest. 

\section{Results: Superintermolecular orbitals}

\begin{figure}%[htbp]
\centerline{\includegraphics[angle=0,width=1\columnwidth]{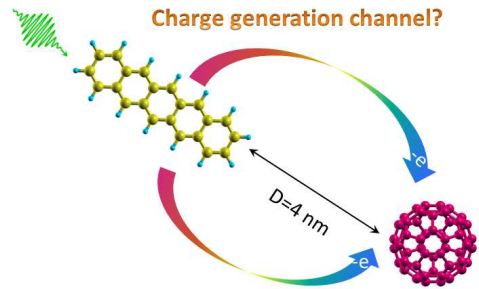}}  
\caption{Light first strikes \ce/pentacene complex and creates a
  singlet, followed by singlet fission into triplets. But electrons
  and holes have to dissociate from each other to become free charge
  carriers. The central question is what the channel states are for
  charge generation in organic solar cells. We show that the
  superintermolecular orbitals may offer an answer.  }
\label{fig1}

\end{figure}

Photovoltaic effects depend on an efficient charge transfer from a
donor (D) to an acceptor (A) \cite{jpc95}. Figure \ref{fig1}
schematically illustrates that light first strikes \ce/Pc but the
subsequent charge transfer relies on channel states. While there is no
detail study of these channel states, several studies have estimated
the size of channel states to be around 3-4 nm
\cite{gelinas,barker,bernardo,heiber,dutton,davino}. This distance
corresponds to a binding energy of 0.1 eV, which can be reasonably
approximated as $E_B=q^2/4\pi\epsilon r_{CT}$. Here $q$ is the charge,
$r_{CT}$ is the separation between average electrons and holes in the
parent charge-transfer (CT) state.  However, no ordinary molecular
orbitals can be as big as 4 nm.  Being unaware of possible relevance
to photovoltaics in \ce/Pc, Feng and her coworkers \cite{feng}
reported some very peculiar molecular orbitals in \ce, resembling the
atomic orbitals, but with a much larger radius. They are not localized
around the atoms of the cluster, but rather they belong to the entire
cluster, for which they called superatomic molecular orbitals, or
SAMOs. They detected these SAMOs using the scanning tunneling
microscope (STM), where the voltage bias is gradually tuned. Their
appearance is due to the partial delocality of outer shells of carbon
atoms which jointly create a potential. Such a potential allows
electrons to partially delocalize around the entire molecule.  These
orbitals have a distinctive shell structure from $1s$ up to $1d$ and
are optically dark states.  When we were investigating
SAMO\cite{ijmp2015}, we were keenly aware of the large size of those
SAMOs.  We notice that the $1s$ orbital has a size close to \cm
\cite{ijmp2015}.

To begin with, we employ Gaussian09 \cite{gaussian} to separately
optimize \ce, pentacene and \ce/Pc structures. We use the Becke,
3-parameter, Lee-Yang-Parr (B3LYP) method and a correlation-consistent
polarized valence double-zeta (cc-pVDZ) basis. The results are fully
consistent with our and other previous calculations
\cite{pra11,johansson12} in both the eigenenergies and wavefunctions.
The optimized coordinates in Gaussian09, without further optimization,
are used as an input for Octopus \cite{octopus}. The reason is that
Octopus uses grid mesh and slightly breaks the symmetry of degenerate
eigenstates. Although the change in energy is small, we worry that the
introduced force may be too great if we use it to optimize our \ce/Pc
complex.

 As done by many researchers \cite{Yi,Yang,li,ryno}, we
consider both edge-on and face-on configurations.  In the edge-on
configuration one end of Pc aims at the hexagons/pentagons on \ce,
while in the face-on configuration, the plane of Pc faces the
hexagons/pentagons on \ce.

 \begin{widetext} 
\begin{center}
\begin{table}
\caption{Coulomb and exchange matrix elements (in units of eV) among
  LUMO (from 172 to 174) and LUMO+1 (175). }
\begin{tabular}{c|ccccllll}
\hline\hline
~~\hspace{0.5cm}&\multicolumn{4}{l}{$K(nm|mn)$ (eV)}& \multicolumn{4}{l}{$J(nm|nm)$ (eV)}\\
\hline
$n$\textbackslash$m$   &172  & 173   & 174     &175 \hspace{1cm} &172 & 173 &174 &175
\\
\hline
172     &3.64 & 3.42  &  3.42   &0.92  \hspace{1cm} &-- & 0.107
&0.107& 0.555$\times 10^{-6}$  \\
173     & 3.42    & 3.64 &3.42 &0.93 \hspace{1cm} &0.107 & -- &0.108
&0.169$\times 10^{-5}$ \\
174     &3.42 & 3.42 &3.64 &0.92 \hspace{1cm} &0.107 & 0.018 &-- &0.916$\times 10^{-6}$ \\
175     &0.92 & 0.93 &0.92 &4.51 \hspace{1cm} &0.555$\times 10^{-6}$
&0.169$\times 10^{-5}$ &0.916$\times 10^{-6}$ &-- \\
\hline
\hline
\end{tabular}
\label{table1}
\end{table}
\end{center}
 \end{widetext}

\subsection{Edge-on}

\begin{figure}[tb]
%\includegraphics[angle=0,width=0.3\columnwidth]{wf-st0205x.eps}   
%\includegraphics[angle=0,width=0.3\columnwidth]{wf-st0205z.eps}    
%\includegraphics[angle=0,width=0.3\columnwidth]{wf-st0205y.eps}   
%\includegraphics[angle=0,width=0.3\columnwidth]{wf-st0205x0.eps}   
%\includegraphics[angle=0,width=0.3\columnwidth]{wf-st0205y0.eps}    
%\includegraphics[angle=0,width=0.3\columnwidth]{wf-st0205z0.eps}   
%\centerline{\includegraphics[angle=0,width=0.8\columnwidth]{simo4.eps}}   
%\centerline{\includegraphics[angle=0,width=0.8\columnwidth]{pp5.eps}}   
\centerline{\includegraphics[angle=0,width=0.8\columnwidth]{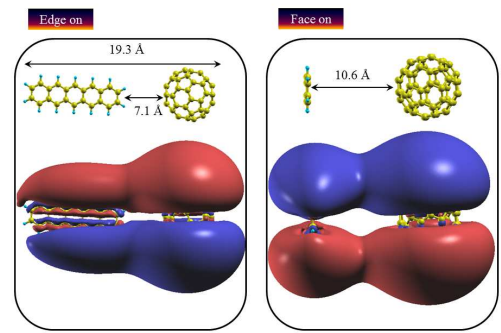}}   
%\centerline{\includegraphics[angle=0,width=1\columnwidth]{chargedsingle.eps}}   
%\centerline{\includegraphics[angle=0,width=1\columnwidth]{c60pc.eps}}   
\caption{ Superintermolecular orbitals in \ce/pentacene for the
  edge-on configuration (left) and face-on configuration (right).  We
  show one representative $1p$ SIMO for each configuration.  $1p$ SIMO
  has orbital number 205. }
\label{simo}
\end{figure}

We start with the edge-on geometry. The distance between the frontier
carbon atoms of Pc and the hexagons on \cm is 7.1 $\rm \AA$, larger
than previous investigations \cite{minami,Yi}. The distance between
the far-left carbon atoms on Pc and the far-right carbon atoms on \cm
is 19.3 $\rm \AA$.  The left figure of Fig. (\ref{simo}) shows one
example for the edge-on configuration. This is the wave function
$\psi_{205}({\bf r})$ for orbital 205 plotted at isovalue of 0.005$\rm
\sqrt[-3]{\rm\AA}$. The color difference denotes the sign of
$\psi_{205}({\bf r})$. Different from SAMOs \cite{feng}, this
superorbital covers both Pc and \cm molecules, or superintermolecular
orbital, SIMO for short. In the language of SAMO, this could be $1p$
SIMO, but for SIMOs, the orbital character is approximate due to the
symmetry reduction. We find that in general SIMOs have special
orientations just as an ordinary molecular orbital. In some cases,
SIMOs are more like SAMOs on an isolated \ce. This spatial preference
is crucial since it allows the electrons to transfer from Pc to \cm
unidirectionally. One special feature, which is inherent from SAMOs,
is that the dipole transition matrix elements between SIMOs and
ordinary molecular orbitals are very small. For this reason, we do not
expect that an optically induced charge transfer occurs from ordinary
molecular orbitals to SIMOs. Instead, the initial bound exciton must
dissociate into SIMOs through tunneling. We recall that Feng \et
\cite{feng} detected SAMOs using STM. The quantum tunneling is also
closer to what happens in solar cells.

However, in photovoltaics, electrons are first excited into those low
lying lowest unoccupied molecular orbitals (LUMOs), which have been
the focus of recent investigations.  For  free charge
generation, majority of theoretical studies start from an initial
state $\phi_i^m({\bf r})$ localized on D, where $m$ is the
multiplicity of state $i$ and ${\bf r}$ is the electron coordinate.
One hopes that this initial state ends up to a final state
$\phi_f^n({\bf r})$ localized on A. In the many-body
picture \cite{aki}, one often starts from configurations like \be
|\Psi\rangle=a |\phi_i^m({\bf r}_1) \phi_f^n({\bf r}_2) \rangle + {\rm
  high-order~ terms.} \ee If $|\phi_i^m({\bf r}_1) \phi_f^n({\bf
  r}_2)\rangle$ takes a significant weight on the many-body
wavefunction, so CT is realized. This idea is simple and attractive,
but faces a dilemma.  To have a large contribution from configuration
$|\phi_i^m({\bf r}_1) \phi_f^n({\bf r}_2)\rangle$, the Coulomb and
exchange interaction matrix elements must be large, but this leads to
a large binding energy, detrimental to free charge carrier
generation \cite{jail}. On the other hand, if the above elements are
small, then the coupling is weak and the transition to CT states is
less likely.  We compute all the Coulomb and exchange integrals from
the lowest unoccupied molecular orbital (LUMO) up to LUMO+1; there are
in total four orbitals since LUMOs are nearly degenerate. Table I
shows all the Coulomb and exchange integrals among all the LUMO
states.  We find that the strongest interaction is 4.51 eV, which is
more than forty times larger than the disorder energy of 0.1 eV
estimated by Clarke and Durrant \cite{clarke}. This leads to a high
binding energy for excitons, thus it is detrimental to free charge
carrier generation \cite{jail}. This simple estimate highlights that
initially excited states by the light are unlikely the same states
that are responsible for final charge transfer and charge
separation. A different group of states engage the final step of
charge generations.

\begin{figure}
\centerline{\includegraphics[angle=270,width=0.8\columnwidth]{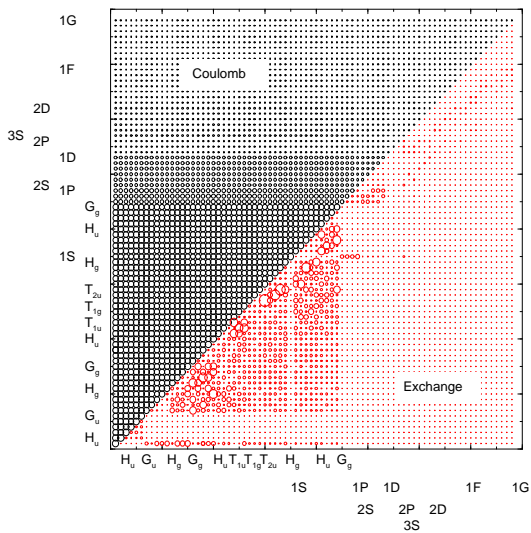}}
\caption{Coulomb and exchange matrix elements between pairs of states
  in native \ce. There are 3081 elements.  The magnitude of matrix
  elements is proportional to the radius of circles, and all the
  Coulomb elements are rescaled by multiplying 0.15 .  Since the
  matrix is symmetric, the upper triangle shows the Coulomb integral,
  while the lower triangle shows the exchange integral.  }
%\label{fig3}
\label{coulombc60}
%\label{fsamo}
\end{figure}

The above calculation is only limited to four unoccupied
orbitals. Before we present results for \ce/Pc, we decide to
completely map out all the matrix elements for all the states from the
highest occupied molecular orbital (HOMO)-4 through $1g$ SAMOs in
native \ce. There are 3081 Coulomb and exchange integrals. All the
calculations are carried out at Berkeley National Laboratory's
National Energy Research Computing Center.  Figure \ref{coulombc60}
shows a complete list of those matrix elements. Since these matrix
elements are symmetric with respect to the state permutation (other
combinations have a much small amplitude, thus not shown), we only
show the upper triangle for the Coulomb integral and the lower
triangle for the exchange integral.  Both the horizontal and vertical
axes denote the states. The SAMOs state labels are slightly off the
axis for clarity. Along the horizontal axis, the second $H_u$ state
from the left is our HOMO, and the first $T_{1u}$ is our LUMO.  The
radii of the circles are proportional to the magnitude of
integral. The Coulomb integrals are in general much larger, so when we
plot them, we reduce their size by multiplying them by 0.15. A general
pattern emerges. For ordinary molecular orbitals, the Coulomb and
exchange integrals are much larger.  The exchange integrals are much
less uniform than the Coulomb integral, since the former greatly
depends on the phases of the wavefunctions.  SAMOs' integrals are also
sizable, in particular for $1s$ SAMO, but once we are above $1p$ SAMO,
both Coulomb and exchange integrals drop very quickly. This opens a
door for delocalized and weak-interacting SAMOs to participate charge
generating process.

\begin{figure}
\centerline{\includegraphics[angle=0,width=0.8\columnwidth]{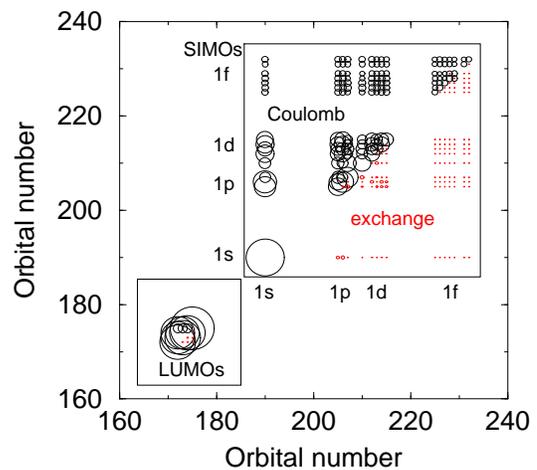}}
\caption{Coulomb and exchange matrix elements between pairs of
  states. The magnitude of matrix elements is proportional to the
  radius of circles.  Since the matrix is symmetric, the upper-left
  triangle shows the Coulomb integral, while the lower-right triangle
  shows the exchange integral.  The largest Coulomb integral (4.51 eV)
  is between LUMO+1 (orbital number 175) localized on Pc; and the
  smallest (0.43 eV) is among $1f$ SIMOs (orbital numbers 226 and
  228).  The exchange integrals are in red small circles and extremely
  small.  }
%\label{fig3}
\label{coulomb}
%\label{fsamo}
\end{figure}

To build a case for SIMOs, we also compute the Coulomb and exchange
integrals and we find that similar to SAMOs, their values are an order
of magnitude smaller. Figure \ref{coulomb} compares the Coulomb (upper
triangle) and exchange (lower triangle) integrals for LUMOs and SIMOs.
The largest circle represents 4.51 eV (which is between LUMO+1). The
Coulomb interaction drops quickly once we are above the $1s$ SIMO. The
smallest Coulomb interaction is for $1f$ SIMOs, only 0.43 eV. If we
consider the dielectric constant of the medium about 3, this
interaction is reduced to 0.14 eV, very close to the disorder
energy. These small Coulomb and exchange integrals are also reflected
in the small transfer integral used by Smith and Chin \cite{smith}.
They concluded that the transfer integrals are no larger than 8 meV,
extremely tiny in comparison to those in \cm \cite{prb94}.

In 2016, in poly(3-hexylthiophene)/fullerene blends, D'Avino
\et\cite{davino} argued that the bound localized charge-transfer (LCT)
states coexist with delocalized space-separated states because LCT
states hybridize with singlets. In a later study \cite{davino2}, they
also suggested that both \cm and its derivative may sustain
high-energy states that spread over a few tens of molecules by
pointing out sizable intermolecular delocalization of the electron
wavefunction. Here our SIMOs present an alternative.

\subsection{Face-on}

We also consider the face-on geometry. In this configuration, the
distance between Pc and \cm is 10.6 $\rm \AA$, also larger than many
prior studies. This configuration is considered to be the most
favorable one for the charge transfer and charge separation. The right
figure of Fig. \ref{simo} shows $1p$ SIMO. It is interesting that
although the face-on geometry is so different from the edge-on
geometry, the SIMO retains its shape well. The wavefunction on Pc has
a larger amplitude than that for the edge-on configuration. This may
explain why it is more efficient, since the orbital is very
delocalized. This meets one of the requirements for the channel
states. Therefore, once an electron tunnels into this orbital, it has
an excellent chance to transfer to \ce.

\begin{figure}
\centerline{\includegraphics[angle=0,width=0.8\columnwidth]{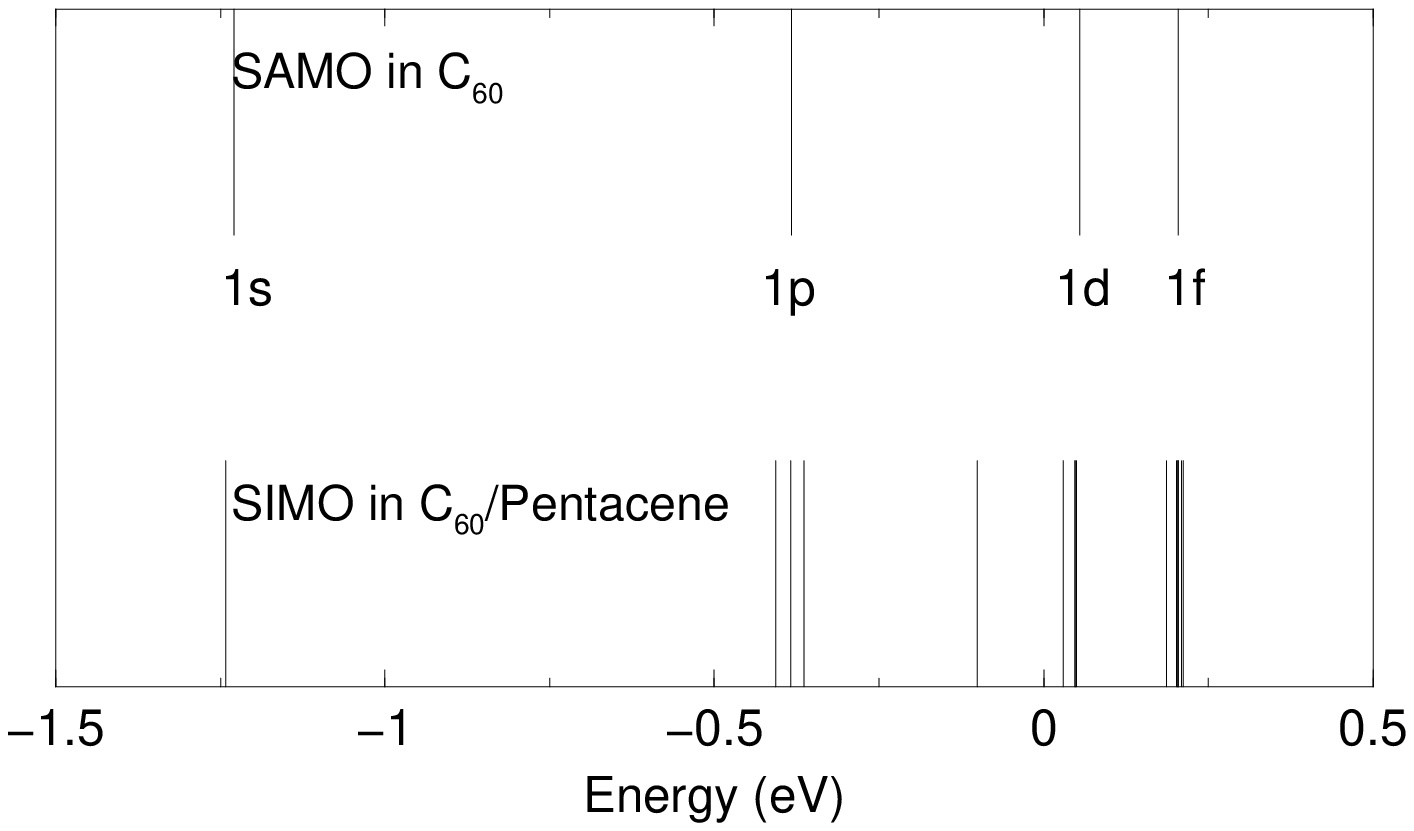}}

{\includegraphics[angle=0,width=0.3\columnwidth]{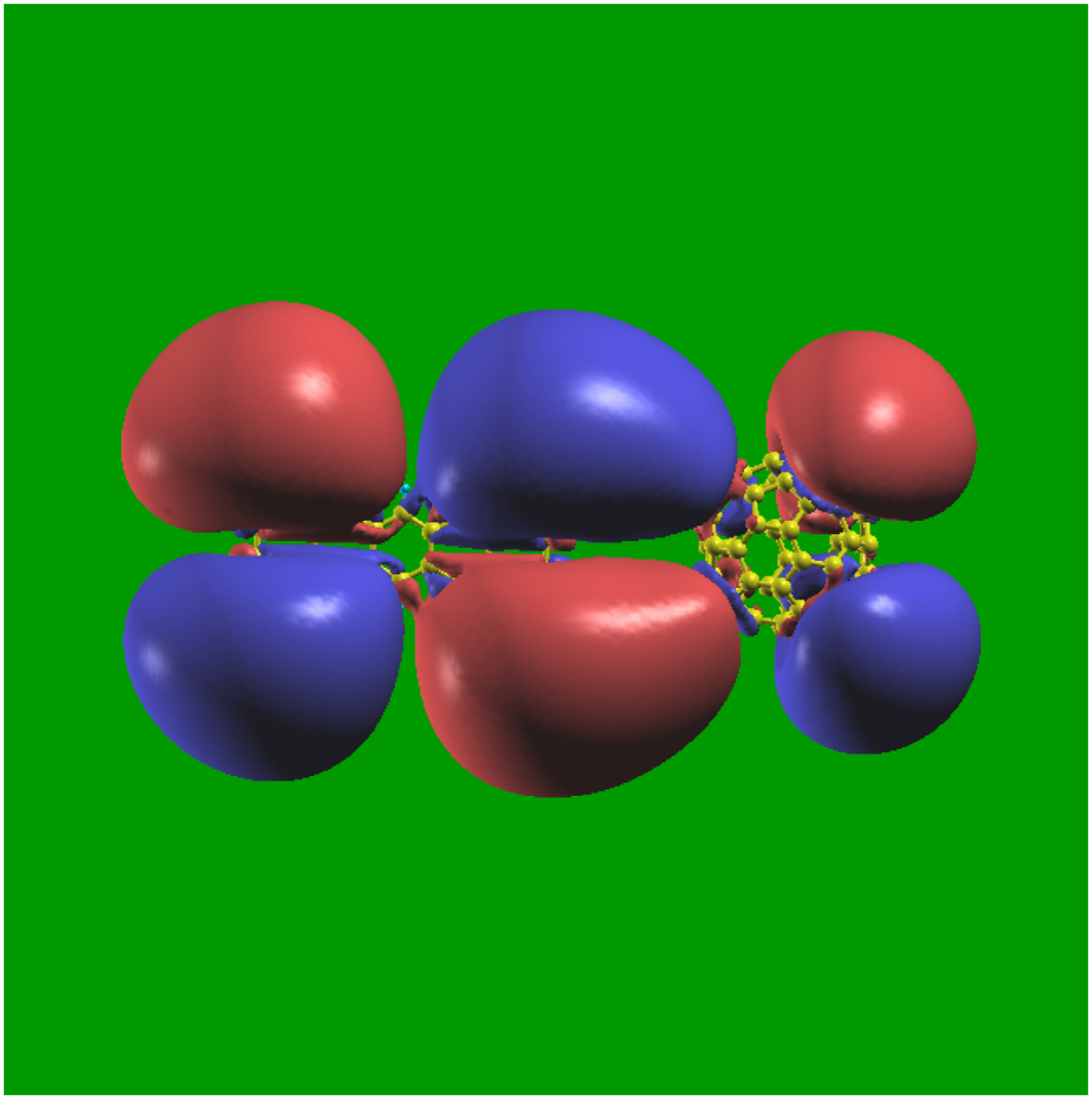}}
{\includegraphics[angle=0,width=0.3\columnwidth]{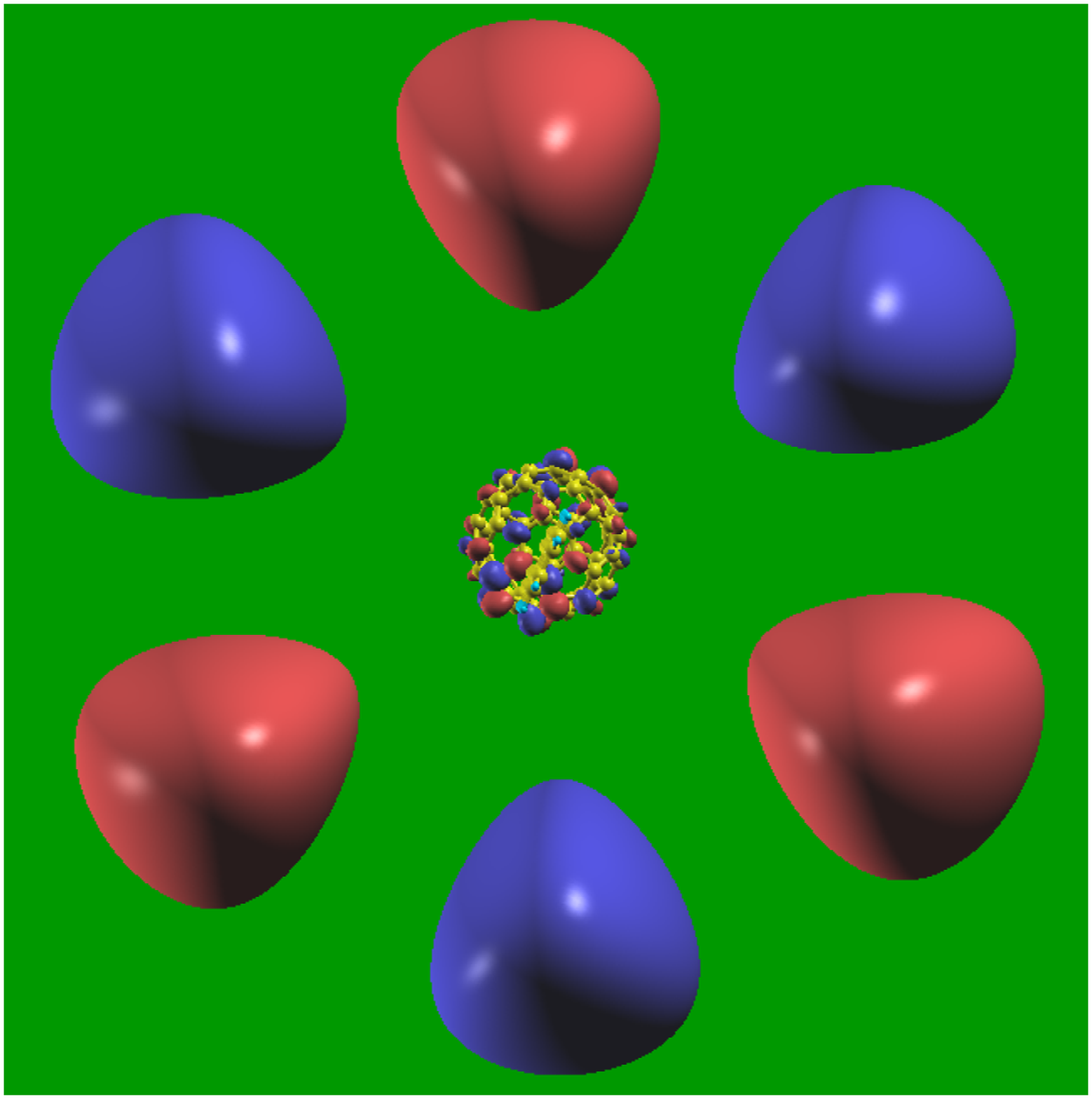}}
\caption{ (Top panel) Energy level comparison between SIMOs of
  \ce/pentacene and SAMOs of \ce. The energy in $1s$ states is
  similar. A small splitting is noticed in $1p$ state. $1d$ state has
  the largest shift of 0.16 eV, due to an overlap in the
  $d$ orbitals.  Change in $1f$ is small. (Bottom panel) Left:
  Wavefunction of $1d$ SIMO; Right: Wavefunction of $1f$ SIMO. }
%\label{fig3}
\label{energy}
%\label{fsamo}
\end{figure}

Energetically, the SIMOs appear in the same energy window as native
SAMOs in \ce. Naturally, this also depends on the spatial orientation
of Pc and \ce. Figure \ref{energy} (top panel) compares the SIMO
energies in \ce/Pc with those of SAMOs in \ce. We see that $1s$ SIMO
and $1s$ SAMO are aligned with each other. However, $1p$ SAMO is now
split into three nearly degenerate levels, due to the symmetry
reductions as explained above. The difference becomes bigger for $1d$
SIMOs. We see that the lowest $1d$ SIMO is shifted down by 0.16 eV.
We understand why it is so. The lower left figure shows that the
orbital wavefunction is delocalized over Pc and \cm and there are
nodal lines in between. The lobes of the orbital overlaps
strongly. This lowers the orbital energy significantly. By contrast,
other orbitals do not change too much. $1f$ SAMO is also split. To
show the orbital 228 has a $f$ character, we rotate the complex
structure so pentacene points out of the page. It is clear that the
orbital retains the $f$ character, but the orbital is elongated along
the Pc-\cm axis. Quantitatively, $1f$ SAMO in native \cm is at 0.204
eV, while $1f$ SIMOs lie between 0.186 and 0.212 eV.  Interestingly,
Pavlyukha and Berakdar \cite{Pavlyukha} showed that SAMOs are
long-lived, coincident with the experimental observation \cite{chep}.
This is what a channel state is supposed to be.  These agreements
constitute strong evidence that the SIMOs may serve a possible channel
for charge separation. However, it is difficult to detect these
delocalized excited states \cite{rao} optically since it may have a
low absorption cross section \cite{osterloh}. One possible method to
test our prediction is the transport measurement. Such a measurement
is in fact more directly related to charge transfer in photovoltaics
than the optical means.  Finally, we notice that there is an ongoing
debate how or whether hot charge-transfer excitons assist free carrier
generation \cite{nan,shen}. But even if hot CT excitons do play a role
\cite{jail}, the binding energy of interfacial CT exciton after
initial excitation is too high for the thermal activation to climb out
of the Coulomb trap.  We argue that SIMOs reported here may provide an
alternative path to CT.

\section{Conclusions}

We have carried out the first-principles density functional
calculation in \ce/Pc complex and find that there exist a group of big
superintermolecular orbitals. These orbitals are very delocalized and
cover both Pc and \ce. We find that SIMOs have the right spatial and
energetic characters to channel the initially bound electrons and
holes into free charge carriers.  Spatially, they are much larger than
ordinary molecular orbitals, close to 1 nm, a critical distance for
CT. They have a clear spatial orientation from Pc to \ce, a crucial
element that greatly facilitates triplet dissociation into free charge
carriers.  Energetically, both exchange and Coulomb interactions of
SIMOs are very small, on the order of 0.1 eV, a value that matches the
Clarke-Durrant disorder energy of 0.1 eV \cite{clarke}. Thus, our
finding highlights an unexpected benefit from SIMOs and points a
possible strategy for tailoring material properties toward
high-efficient organic solar cells. One possible method to enhance
charge generation is to employ a larger fullerene, where SIMOs can be
made even larger. This is consistent with the experiment \cite{chen},
where they showed that large fullerene crystals can enhance charge
separation yields. We expect that our finding will motivate further
experimental and theoretical investigations on these exciting
opportunities at the frontier of photovoltaics.

%\vspace{0.5cm}

%{\sf Acknowledgments }\\ 
\acknowledgments This work was supported by the U.S. Department of
Energy under Contract No.  DE-FG02-06ER46304 (GPZ).  This research
used resources of the National Energy Research Scientific Computing
Center, which is supported by the Office of Science of the U.S.
Department of Energy under Contract No. DE-AC02-05CH11231.

 $^*$gpzhang@indstate.edu\\

\appendix

\section{Calculation of Coulomb and exchange integrals}

In this appendix, we explain how the Coulomb and Exchange integrations
are done using combination of MPI and OpenMP parallelization. The
Coulomb integral is 
\begin{widetext}
\be
K(nm|mn)=\frac{1}{4\pi\epsilon_0}\int_{-\infty}^{\infty}\int_{-\infty}^{\infty}\int_{-\infty}^{\infty}
d{\bf r}_1d{\bf r}_2 \psi_{n}^\dagger({\bf r}_1) \psi_{m}^\dagger({\bf
  r}_2) \frac{e^2}{r_{12}} \psi_{m}({\bf r}_2)\psi_{n}({\bf r}_1), \ee
and exchange integral \be
J(nm|nm)=\frac{1}{4\pi\epsilon_0}\int_{-\infty}^{\infty}\int_{-\infty}^{\infty}\int_{-\infty}^{\infty}d{\bf
  r}_1d{\bf r}_2 \psi_{n}^\dagger({\bf r}_1) \psi_{m}^\dagger({\bf
  r}_2) \frac{e^2}{r_{12}} \psi_{n}({\bf r}_2)\psi_{m}({\bf r}_1), \ee \end{widetext}
where $\psi_n$ and $\psi_m$ are the respective wavefunctions for state
$n$ and $m$, $r_{12}$ is the distance between two electrons situated
at positions ${\bf r}_1$ and ${\bf r}_2$, and $e$ is the charge
unit. Although we may use the medium permittivity in the above two
equations, we decide to use the permittivity in vacuum $\epsilon_0$ so
the reader can verify our results easily. Note that we only consider
the paired states, since other forms have a much smaller integral.
The integral over ${\bf r}_2$ is parallelized using OpenMP, and
distributed evenly to processors in each MPI task. The integral over
${\bf r}_1$ is parallelized using MPI. The final results in each MPI
task is summed up using MPI reduction. The only serial part of the
implementation is the file IO and input of the wavefunctions.  Since
the Coulomb and exchange integrals have singularity, the treatment
needs some caution although their overall contributions are
small. Around the singularity, we replace the cube (grid mesh used in
Octopus) by a sphere. The spherical coordinate allows an analytic
integration. Then we rescale the volume of the sphere to that of a
cube. Finally, we add the integral back to the final sum. This method
is very accurate.

The above implementation is post-processed in our own code, not in
Octopus.  The hybrid MPI/OpenMP calculation is set up according to the
computing system hardware structure. The supercomputer system Cori at
National Energy Research and Scientific Computing Center (NERSC) at
Berkeley National Laboratory has dual CPUs with a total of 32 cores
per node. With this system we run 4 MPI tasks and 8 OpenMP threads on
each node to obtain optimal performance.

\newcommand{\comment}{}

%/home/gpzhang/c60/samo/pentacene/vh3

%\begin{figure}
%\centerline{\includegraphics[angle=270,width=.8\columnwidth]{fig4.eps}}
%\caption{
%Optical absorption spectrum (a) without  and (b) 
%with superatomic molecular orbital
%contribution. The figure is in the same arbitrary scale. 
%(c) Difference spectrum with and without SAMO contributions. A peak
%appears around 9 eV. This is due to the transition between the $G_u$
%and $1d$ SAMO. 
%  }
%\label{fig4}
%\end{figure}

\clearpage

\end{document}